\documentclass[aps,prd,reprint,groupedaddress,showpacs]{revtex4}

\usepackage{graphicx}
\usepackage{dcolumn}
\usepackage{bm}

\newcommand{\pom}{\tt I\! P}

\begin{document}

\title{Photoproduction of  pentaquark states at the LHC}

\author{V. P. Gon\c calves}

\email[]{barros@ufpel.edu.br}

\affiliation{High and Medium Energy Group, \\ Instituto de F\'{\i}sica e Matem\'atica,  Universidade Federal de Pelotas (UFPel)\\
Caixa Postal 354,  96010-900, Pelotas, RS, Brazil.}

\author{M. M. Jaime}

\email[]{mmedinaja@gmail.com}

\affiliation{Universidad ECCI \\ Cra. 19 No. 49-20, C\'odigo Postal 111311, Bogot\'a, Colombia}

\begin{abstract}
In this paper we investigate the production of pentaquark states in the photon -- proton interactions present in hadronic collisions at the RHIC and LHC. We consider two phenomenological models for the $J/\Psi$ photoproduction that consider  
the presence of the $P_c(4312)$, $P_c(4440)$ and $P_c(4457)$ resonances in the $s$ -- channel of the $\gamma p \rightarrow J/\Psi p$ reaction. The rapidity distribution is estimated for $pA$ collisions at the RHIC and LHC. Predictions for $pPb$, $Pb \, Ar$ and $Pb \,He$ fixed -- target collisions at the LHC are also presented. We demonstrate that the  experimental analysis of the $J/\Psi$ photoproduction in fixed -- target collisions  can provide complementary and independent checks on these states, and help to understand their underlying nature.
\end{abstract}


\maketitle


In the last  years a series of  charmoniumlike exotic states has been announced at various experimental facilities (For reviews see, {\it e.g.}, Refs. 
\cite{Olsen:2017bmm,Liu:2019zoy}). 
  Such states  decay to final states that contain a charm and an anticharm but cannot be easily accommodated in the remaining unfilled states in the $c\bar{c}$ level scheme.
Recently, the LHCb Collaboration has reported their new results for the $\Lambda_b^0 \rightarrow J/\Psi p K^-$ decay, which indicate the existence of three narrow pentaquark states: $P_c(4312)$, $P_c(4440)$ and $P_c(4457)$ \cite{lhcb_penta}. Such observation motivated a series of theoretical studies  as well as the search of these states in other processes (See e.g. Refs.  
\cite{Chen:2019asm,Xiao:2019mst,Chen:2019bip,Cheng:2019obk,Liu:2019tjn,He:2019ify,Guo:2019kdc,gluex_penta,d0_penta}). In particular, the GlueX Collaboration  searched for the pentaquark states in the near-threshold $J/\Psi$ exclusive photoproduction off the proton \cite{gluex_penta} and the D0 Collaboration have analyzed the production of these states in $p\bar{p}$ collisions \cite{d0_penta}. While the experimental results from  D0 Collaboration find an enhancement in the $J/\Psi p$ invariant mass consistent with a sum of the resonances $P_c(4440)$ and $P_c(4457)$ reported by the LHCb Collaboration, the GlueX Collaboration did not see evidence for them. Such results indicate that the proposition of new reaction channels that can be used  to probe  the existence  and to decipher the nature of these states is timely and welcome.

\begin{figure}[t]
\begin{tabular}{cc}
\includegraphics[scale=0.3]{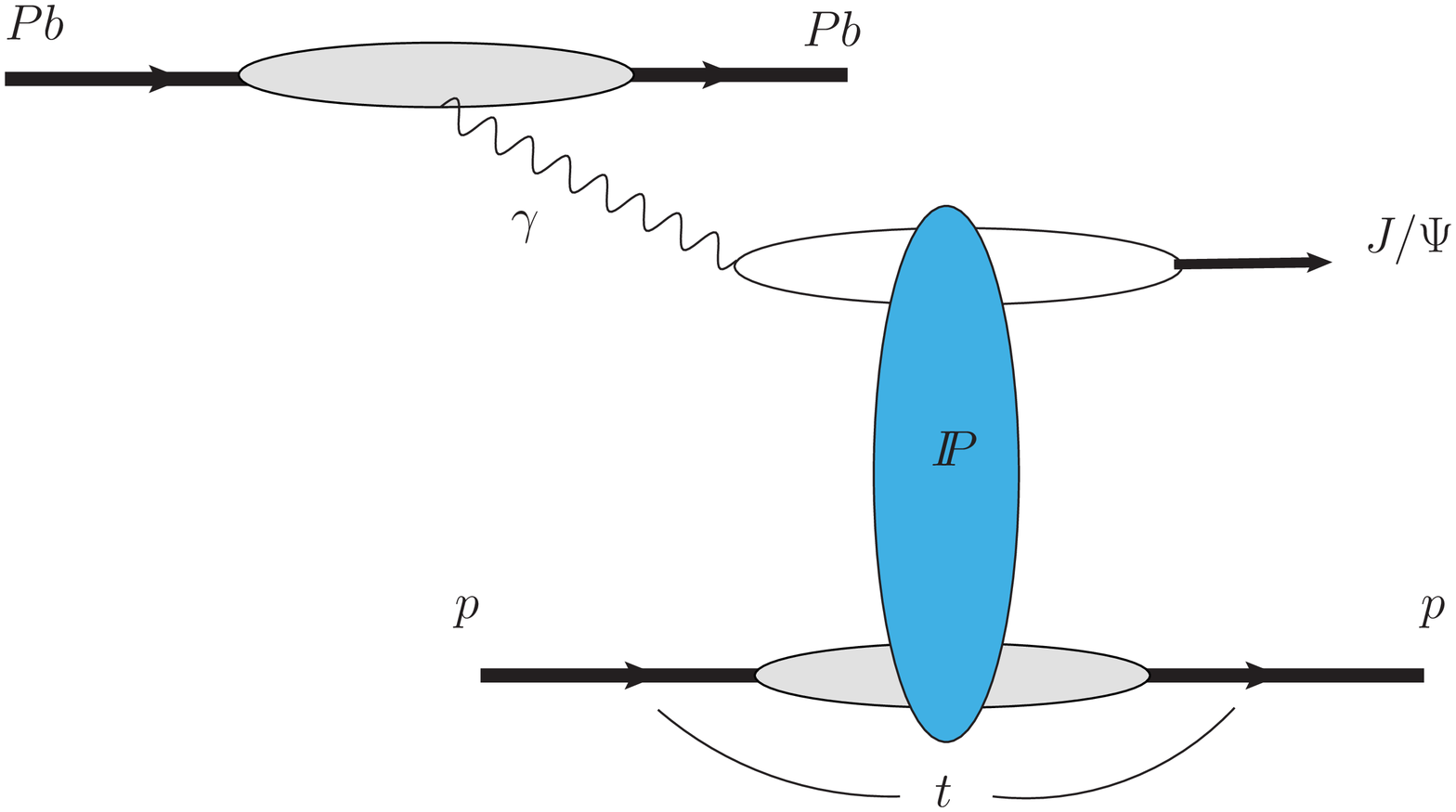} &
\includegraphics[scale=0.3]{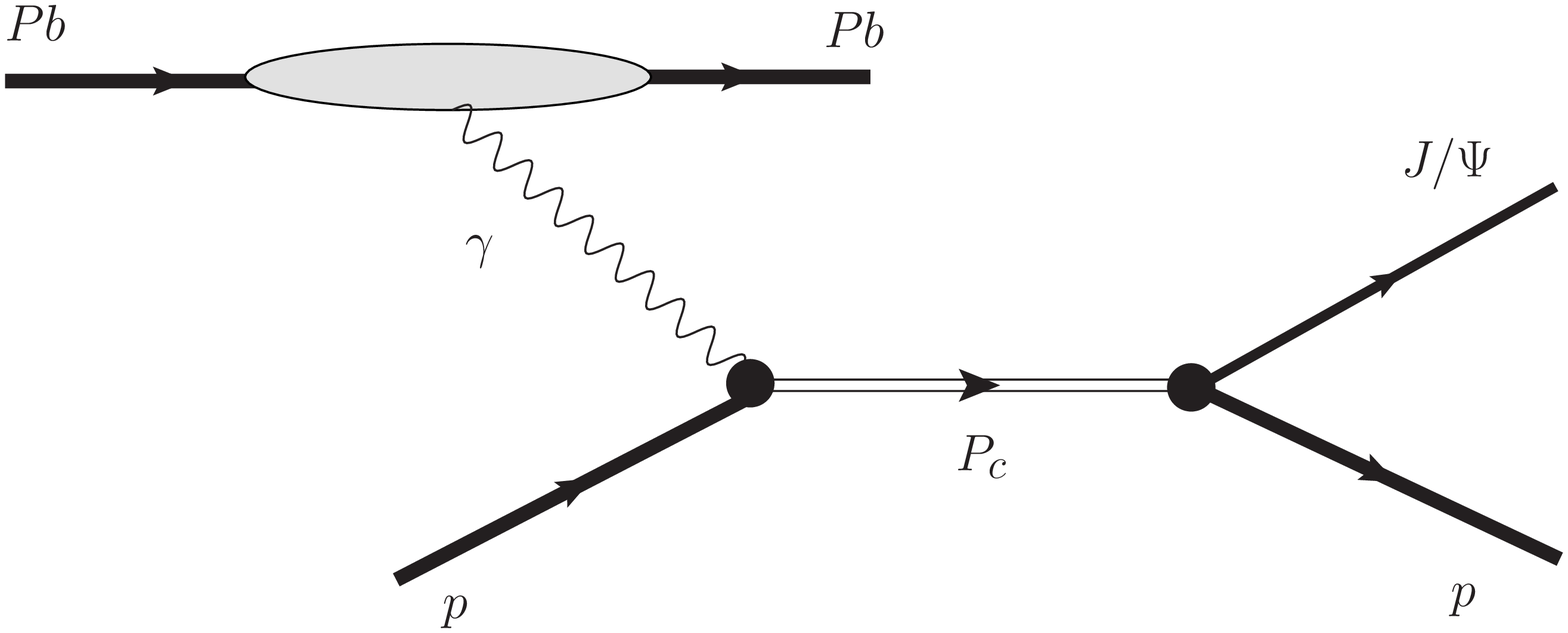} \\
(a) & (b)
\end{tabular}
\caption{ The $Pb + p \rightarrow Pb +  J/\Psi + p$ process through the (a) Pomeron exchange and (b) $P_c$ production. }
\label{Fig:diagrama}
\end{figure}

In this paper we will analyze the possibility to probe the pentaquark states in photon -- induced interactions at the RHIC and LHC, which are dominant at large impact parameters ($b > R_1 + R_2$, where $R_i$ is the hadron radius). The results obtained in Refs. 
\cite{vicdaniel,exotic_vicbert,exotic_bruno} have demonstrated that the properties of exotic states can be studied by considering its production in the photon -- photon interactions present in ultraperipheral collisions. Another possibility is to consider the production of exotic states in photon -- hadron interactions. 
As demonstrated for the first time in Ref. \cite{vicmario} and more recently in Ref. \cite{klein_penta}, hadronic collisions can be used to study the production of charged $Z_c$ charmoniumlike states in photon -- hadron interactions. In our analysis we will extend  previous studies for the exclusive $J/\Psi$ photoproduction in $pPb$ collisions at the LHC,  which  estimated the contribution of the Pomeron exchange for the cross section [represented in Fig. \ref{Fig:diagrama} (a)], by  taking into account  the  contribution associated to the presence of the $P_c(4312)$, $P_c(4440)$ and $P_c(4457)$ resonances in the $s$ -- channel [See Fig. \ref{Fig:diagrama} (b)].  Motivated by the results presented in Ref. \cite{vicmiguel}, where we have demonstrated for the first time that photon --  induced interactions can be studied in fixed -- target collisions at the LHC, we will present our predictions considering $pPb$ collisions in the collider and fixed -- target modes of the LHC. In addition, we also will present predictions for $Pb \, Ar$ and $Pb \, He$ fixed -- target collisions.
In  our analysis we will consider two different models for the description of the pentaquark production in photon -- proton interactions, which allow us to estimate the model dependence of our predictions. 
As we will show below, our results indicate that  the study of the  $J/\Psi$ photoproduction in fixed -- target collisions can be  useful to probe the existence of the pentaquark states.

Initially we present a brief review of the formalism need to describe the $J/\Psi$ photoproduction in $pPb$ collisions. In these collisions, both hadrons act  as a source of 
almost real photons, with the equivalent photon flux associated to the nucleus being enhanced by a factor $Z^2$ in comparison to the proton one. Consequently, in a ultraperipheral $pPb$ collision, where  $b > R_{p} + R_{Pb}$,  the photon-induced interactions become dominant and the total hadronic cross section for the $J/\Psi$ production  can be expressed in terms of the equivalent flux of photons and the photon -- hadron cross section as follows \cite{upc}
\begin{eqnarray}
   \sigma(p + Pb \rightarrow p \otimes J/\Psi \otimes Pb;\,s) =  \int d\omega \,\, n_{p}(\omega) \, \sigma_{\gamma Pb \rightarrow J/\Psi \otimes Pb}\left(W  \right) +  
   \int d \omega \,\, n_{Pb}(\omega)
   \, \sigma_{\gamma p \rightarrow J/\Psi \otimes p}\left(W  \right)\,  \; , 
\label{eq:sigma_pp}
\end{eqnarray}
where $\sqrt{s}$ is center-of-mass energy for the $pPb$ collision, $\otimes$ represents the presence of a rapidity gap in the final state, $\omega$ is the energy of the photon emitted by the hadron and $n_h$ is the equivalent photon flux of the hadron $h$ integrated over the impact parameter. Moreover, $\sigma_{\gamma h \rightarrow J/\Psi  h}$ describes the vector meson production in photon - hadron interactions for a given photon -- hadron center of mass energy $W$, which is given by 
$W = \sqrt{2\,\omega\, \sqrt{s}}$. We will estimate the cross section for ultraperipheral collisions using the STARLight Monte Carlo, in which  the nuclear and proton  photon spectra are calculated as follows \cite{klein,starlight}
\begin{eqnarray}
n_h(\omega) = \int \mbox{d}^{2} {\mathbf b} \, P_{NH} ({\mathbf b}) \,   N_h\left(\omega,{\mathbf b}\right) \,\,, 
\end{eqnarray}
where $P_{NH} ({\mathbf b})$ is the probability of not having a hadronic interaction at impact parameter ${\mathbf b}$ and 
the number of photons per unit area, per unit energy, derived assuming a point-like form factor, is given by 
\begin{equation}
N_h(\omega,{\mathbf b}) = \frac{Z^{2}\alpha_{em}}{\pi^2} \frac{\omega}{\gamma_L^{2}}
\left[K_1^2\,({\zeta}) + \frac{1}{\gamma_L^{2}} \, K_0^2({\zeta}) \right]\,
\label{fluxo}
\end{equation}
with $\zeta \equiv \omega b/\gamma_L$, $\alpha_{em}$ is the electromagnetic fine structure constant,  $\gamma_L$ is the Lorentz gamma factor and $K_0(\zeta)$ and  $K_1(\zeta)$ being the
modified Bessel functions.
In the case of proton -- nucleus collisions, $P_{NH} ({\mathbf b})$ is given by
\begin{eqnarray}
 P_{NH}({\mathbf b}) = \exp\left[
 - \sigma_{nn}  T_{A}({\mathbf b})
 \right].
\end{eqnarray}
with $\sigma_{nn}$ being the total hadronic interaction cross section and $T_{A}$ is the nuclear thickness function, which is calculated from the nuclear density profile, assumed to follow a Woods -- Saxon distribution. In principle, other models can be assumed to describe the photon flux and $ P_{NH}({\mathbf b})$. However, as demonstrated e.g. in Ref. \cite{celsina},  these different treatments modify the behaviour of the flux at large photon energies $\omega$ as well as the cross sections of final states with a large invariant mass. For the  $P_c$ photoproduction, the impact of these different modelling is expected to be small.
The coherent $\gamma Pb \rightarrow J/\Psi Pb$ cross section is expressed in  the STARlight MC as follows
\begin{eqnarray}
\sigma(\gamma Pb \rightarrow J/\Psi Pb) =   \int_{-\infty}^{t_{min}} dt \, \frac{d\sigma(\gamma Pb \rightarrow J/\Psi Pb)}{dt}|_{t=0} \,|F(t)|^2 \,\,
\label{star1}
\end{eqnarray}
where $F(t)$ is the nuclear form factor and $t_{min} = - (M_{J/\Psi}^2/4 \omega \gamma_L)^2$. For Lead ions, the form factor is assumed to be the convolution of a hard sphere potential with a Yukawa potential of range 0.7 fm.   The differential cross section for a photon - nucleus interaction is determined using the optical theorem and the generalized vector dominance model (GVDM) \cite{sakurai}
\begin{eqnarray}
\frac{d\sigma(\gamma Pb \rightarrow J/\Psi Pb)}{dt}|_{t=0} = \frac{\alpha_{em} \sigma^2_{tot}(J/\Psi Pb)}{4 f_{\Psi}^2} \,\,,
\label{gvdm}
\end{eqnarray}
where $f_{\Psi}$ is the $J/\Psi$ - photon coupling and the total cross section for the vector meson - nucleus interactions is found using the classical Glauber calculation. As a consequence, it is possible to express $\sigma(\gamma Pb \rightarrow J/\Psi Pb)$ in terms of  $\sigma(\gamma p \rightarrow J/\Psi p)$ [See Eq. (9) in Ref. \cite{klein}]. Finally, it is important to point out that the formalism can be directly applied to nuclear collisions through the proper modification of   $P_{NH} ({\mathbf b})$ and the nuclear photon fluxes.



\begin{figure}[t]
\includegraphics[scale=0.35]{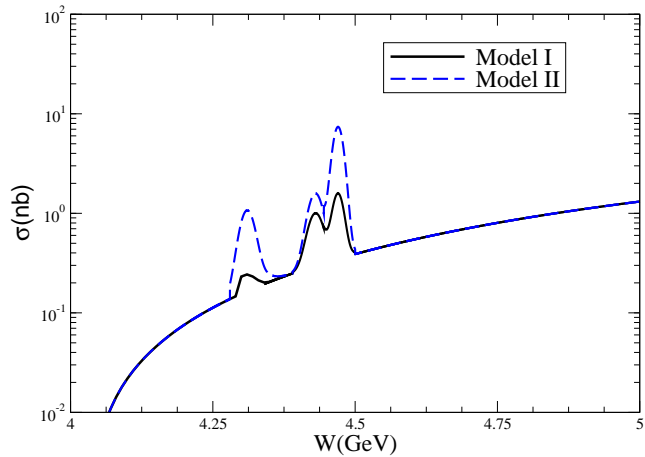} 
\caption{Predictions  for the energy dependence of the $\gamma p \rightarrow J/\Psi p $ cross section taking into account of the Pomeron exchange and the $P_c$ production. The  Pomeron exchange is described by the STARLight MC while the production of the  $P_c(4312)$, $P_c(4440)$ and $P_c(4457)$ resonances  in the $s$ -- channel is described by the phenomenological models proposed in Refs. \cite{model1,model2}. }
\label{Fig:gamap}
\end{figure}

In our analysis we will consider that the $J/\Psi + p$ final state can be generated by a Pomeron exchange and by a $P_c$ resonance in the $s$ -- channel, as represented in the Figs. \ref{Fig:diagrama} (a) and (b), respectively. The contribution associated to the Pomeron exchange for the  $\gamma p  \rightarrow J/\Psi \, p$ cross section will be described following the STARLight MC, where the energy dependence of this cross section is described by a parametrization inspired in the Regge theory given by
\begin{eqnarray}
\sigma_{\gamma p \rightarrow J/\Psi \otimes p} = \sigma_{\pom} \times W^{\epsilon}  \,\,.
\label{sig_gamp}
\end{eqnarray}
The free parameters  on the parametrization, $ \sigma_{\pom}$ and $\epsilon$,  are fitted using the HERA data \cite{hera}, being given by: $ \sigma_{\pom} = 4.06$ nb and  $\epsilon = 0.65$.  In addition, the STARLight MC supplements this cross section by a factor that accounts for its behavior for energies near  the  threshold of production. On the other hand, the contribution associated to the $P_c$ resonance will be described considering two distinct phenomenological models present in the literature \cite{model1,model2}. We will consider the  model presented in the Ref. \cite{model1}, denoted Model I hereafter, in which the cross section for the production of the $P_c$ state in the $s$ -- channel of   the $\gamma p \rightarrow J/\Psi p$ reaction was estimated using the approach proposed in Refs. \cite{volo,rosner}, with the main input being   the branching ratio ${\cal{B}}(P_c \rightarrow J/\Psi p)$. In Ref. \cite{model1}, the authors have constrained the range of values for this quantity
using the branching ratios and fractions measured by the LHCb and GlueX collaborations. In our analysis we will consider the upper values derived in Ref. \cite{model1}, which implies that the associated predictions should be considered an upper limit for the photoproduction of $P_c$ in ultraperipheral collisions. 
In addition,  we also will consider the model proposed in Ref. \cite{model2}, denoted Model II hereafter, where the photoproduction of the $P_c$ states is estimated within the framework of an  effective Lagrangian approach combined with the vector meson dominance assumption \cite{sakurai}. 
In this model, the main inputs in the calculations are the partial decay widths $\Gamma_{P_c \rightarrow J/\Psi p}$ of the different $P_c$ states, which determine the electromagnetic couplings related to the $\gamma p P_c$ vertices and $g_{P_c J/\Psi p}$  coupling constants. Currently, the value of these quantities are still a theme of debate. Following Ref. \cite{model2}, in our calculations using the Model II we will assume that the $J/\Psi p$ channel accounts for $3 \%$ of total widths of the $P_c$ states. 
We refer the reader to the original references \cite{model1,model2} for more details about these phenomenological models for the $P_c$ photoproduction. 
The resulting predictions of these models for the energy dependence of the $\gamma p \rightarrow J/\Psi p $ cross section considering the Pomeron exchange and the production of the  $P_c(4312)$, $P_c(4440)$ and $P_c(4457)$ resonances  in the $s$ -- channel are presented in Fig.  \ref{Fig:gamap}. 
The presence of the resonances modifies the energy dependence of $\sigma(\gamma p \rightarrow J/\Psi p)$ in the range $4.25 \le W \le 4.6$ GeV. We  found that these models differ in its predictions for the impact of the resonances on the cross section, with the Model II predicting a larger contribution associated to the resonances.  For larger and smaller energies, the cross section is dominated by the Pomeron exchange, with our predictions being similar to those derived in the original references \cite{model1,model2}, where different descriptions of the Pomeron exchange were considered. As these distinct models have considered the same set of data to constrain its free parameters, such result is expect.

\begin{figure}[t]
\begin{tabular}{ccc}
\includegraphics[scale=0.45]{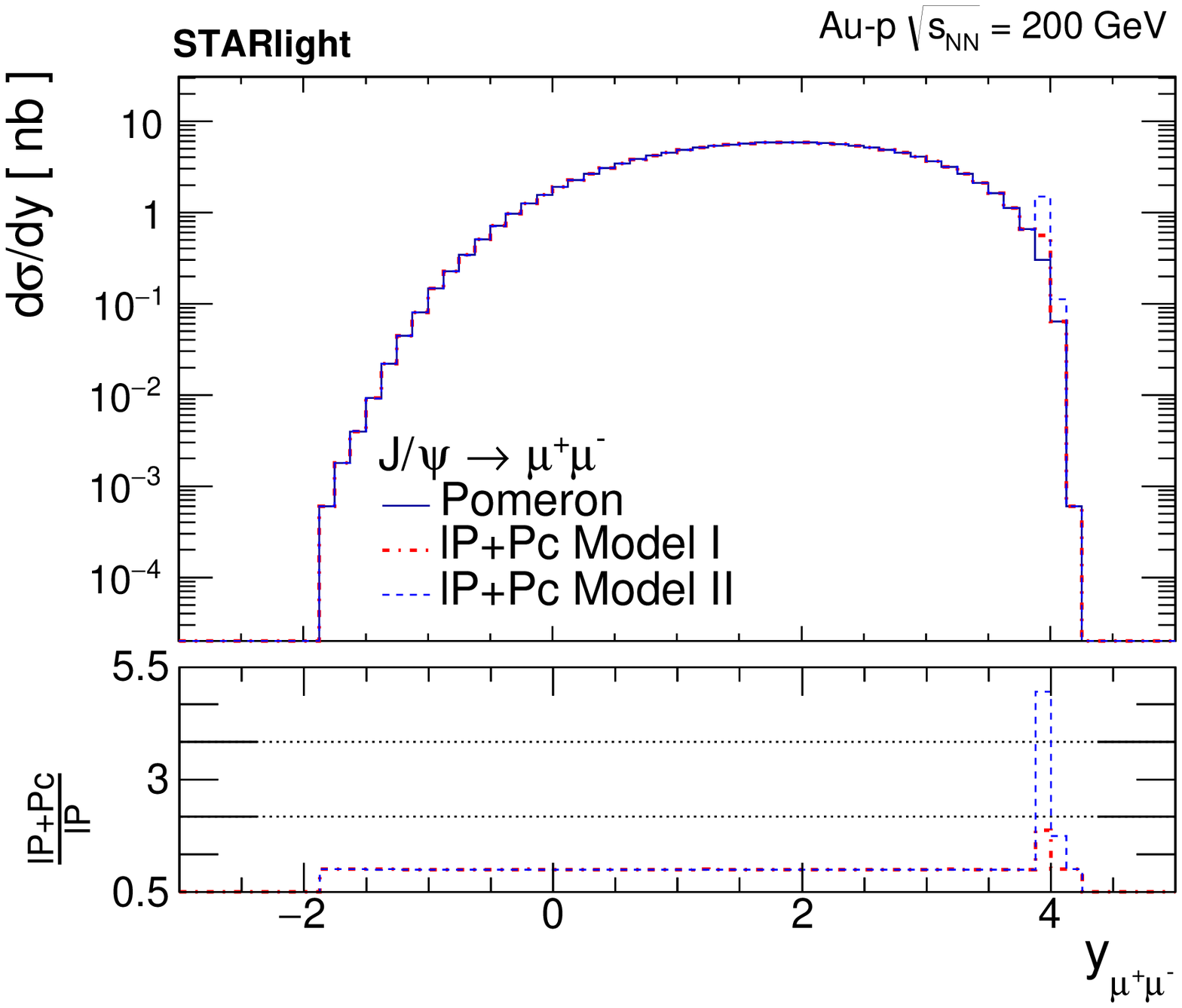} & \hspace{0.2cm} &
\includegraphics[scale=0.45]{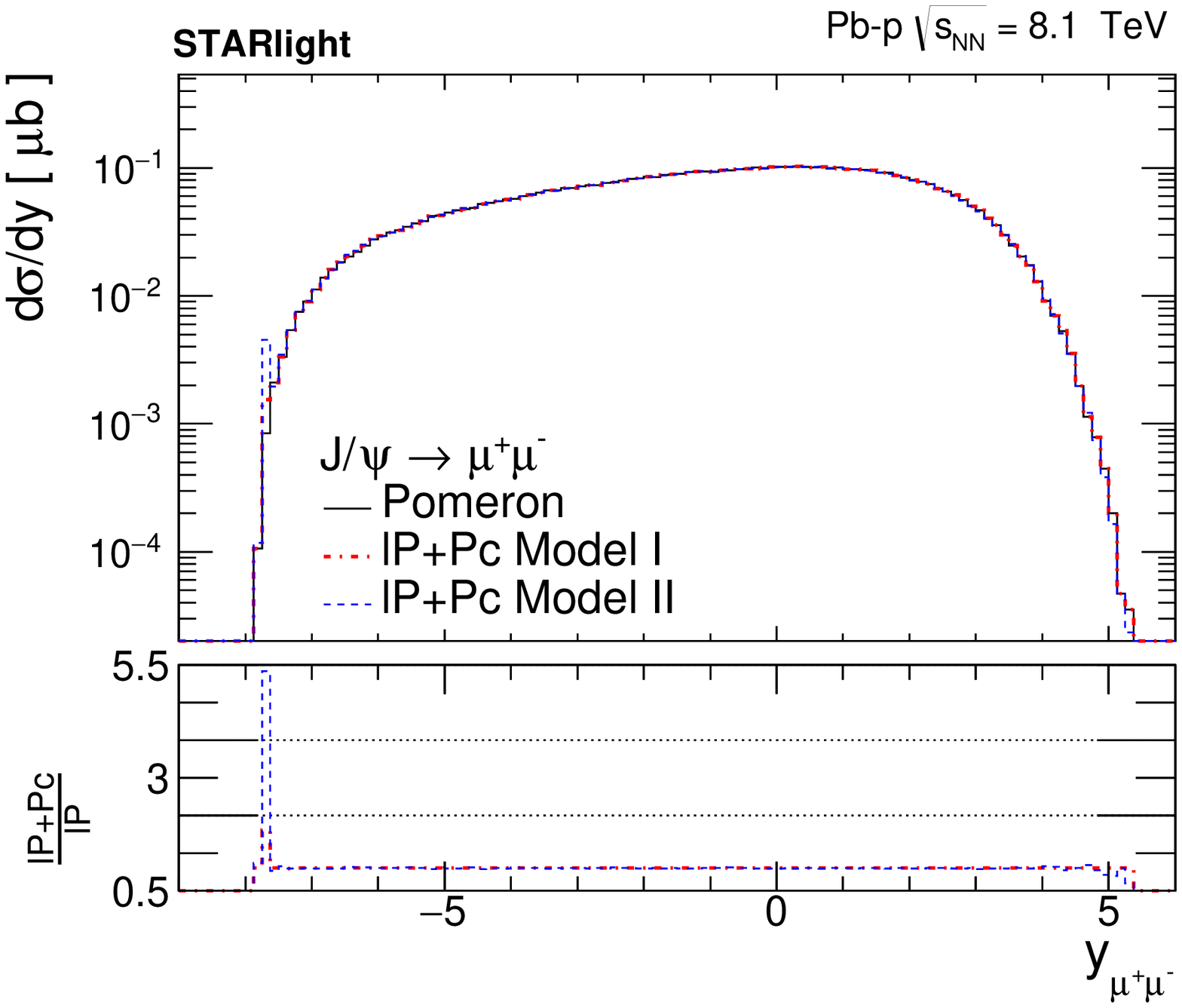} \\
(a) & \,\,\, & (b)
\end{tabular}
\caption{ Rapidity distribution for the $J/\Psi$ photoproduction in (a) $pAu$ collisions at RHIC and (b) $pPb$ collisions at LHC in the collider mode. }
\label{Fig:pentacol}
\end{figure}

In what follows we will consider the phenomenological models discussed above as input in our calculations of the $J/\Psi$ photoproduction in $pA$ collisions.  In order to derive our predictions,   the STARLight MC was modified to  include the contribution of the $P_c$ resonances in the description of the energy dependence of the $\gamma p \rightarrow J/\Psi p $ cross section. 
Initially, let's consider $pA$ collisions in the collider mode for the RHIC and LHC energies. 
For RHIC we will consider $pAu$ collisions at $\sqrt{s} = 0.2$ TeV, while for the LHC we will assume $pPb$ collisions at  $\sqrt{s} = 8.1$ TeV.
 Our predictions for the rapidity distribution of the dimuons generated in the $J/\Psi \rightarrow \mu^+ \mu^-$ decay  are presented in Figs. \ref{Fig:pentacol} (a) and (b) for RHIC and LHC energies, respectively. For comparison we present the predictions derived with  and without the inclusion of the $P_c$ resonances, which are denoted  $\pom + P_c$ and Pomeron, respectively. 
We have that the rapidity distribution is asymmetric about midirapidity ($y_{
\mu^+ \mu^-} \approx 0$) and that the inclusion of the $s$ -- channel contribution implies an enhancement of the rapidity distribution for a given $y_{\mu^+ \mu^-}$, 
 with the magnitude of this enhancement being dependent on the description of the $P_c$ production. Such results are expected. 
For $pA$ collisions the rapidity distributions are asymmetric in rapidity due to the asymmetry on the initial photon fluxes associated to a proton and a nucleus, with the nuclear photon flux being enhanced by a factor $Z^2$.  One important consequence is that  the behaviour of the distribution is dominated  by $\gamma p$ interactions with  the rapidity directly determining the value of the photon energy $\omega$ 
that is being probed: $\omega \propto e^{Y}$. As $W = \sqrt{2\,\omega\, \sqrt{s}}$, we have that for RHIC (LHC) energies, the behaviour of $\sigma(\gamma p \rightarrow J/\Psi p)$ in the energy range  where the presence of the resonances modifies the cross section  is probed for positive (negative) values of rapidity (See Fig. \ref{Fig:gamap}). In the lower part of Fig. \ref{Fig:pentacol}  we present our results for the ratio between the $\pom + P_c$ and Pomeron predictions. We have that the Model I (II) predicts an enhancement of order of 1.8 (5.3), which occurs for $y_{\mu^+ \mu^-} \approx 4.0  \, (- \, 7.5)$ in $pA$ collisions at RHIC (LHC). Unfortunately, this enhancement occurs for rapidities beyond those covered by the current detectors of the RHIC and LHC. Moreover, we only observe one peak, which implies that we will not be able to discriminate the contribution of the different resonances, which are very close in mass.

\begin{figure}[t]
\includegraphics[scale=0.45]{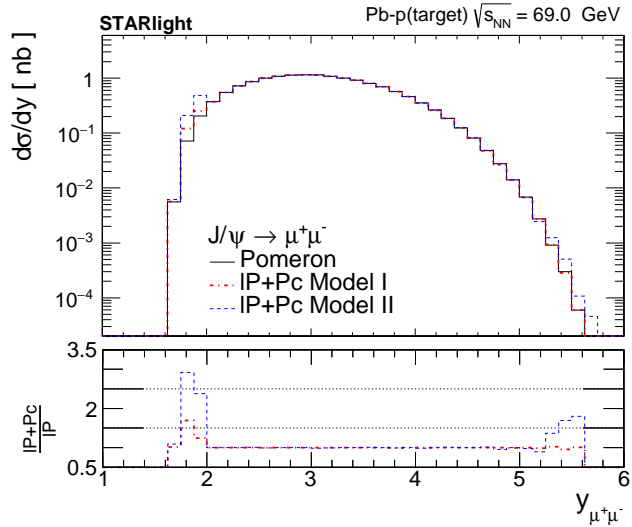} 
\caption{ Rapidity distribution for the $J/\Psi$ photoproduction in  $pPb$ collisions at LHC considering the  fixed -- target mode. }
\label{Fig:penta}
\end{figure}

Let's now investigate the $P_c$ photoproduction in fixed -- target collisions at the LHC. The study of fixed - target collisions at the LHC became recently a reality by the injection of noble gases in the LHC beam pipe by the LHCb Collaboration \cite{lhcfixed} using the System for Measuring Overlap with Gas (SMOG) device \cite{smog}. More results are expected in forthcoming years \cite{after}.  As discussed in detail in Ref. \cite{vicmiguel}, the study of photon - induced interactions  is expected to be possible  in fixed - target collisions. In particular, Ref. \cite{vicmiguel} demonstrated that in fixed -- target collisions we will be able to constrain,  in the kinematical range probed by the LHCb detector,  the behaviour of the $\gamma p \rightarrow J/\Psi p$ cross section at low center -- of -- mass energies, near to the threshold of production. Such results motivate the analysis of the $P_c$ photoproduction in fixed -- target collisions.
In our analysis we will assume $Pb p$ collisions with $\sqrt{s} = 69$ GeV, with the proton being the fixed target. Our results are presented in Fig. \ref{Fig:penta}. For the fixed -- target energy, we have that the rapidity distribution is narrower and concentrated in the rapidity range $1.6 \le y_{
\mu^+ \mu^-} \le 4.6$. The enhancement associated to the $s$ -- channel reaction now occurs for $y_{\mu^+ \mu^-} \approx  \, 1.8$, being of order of 1.6 (3) for the Model I (II). Such result demonstrate that the analysis of fixed -- target collisions at the LHC can be useful to constrain the presence of the pentaquark states. 

\begin{figure}[t]
\begin{tabular}{ccc}
\includegraphics[scale=0.45]{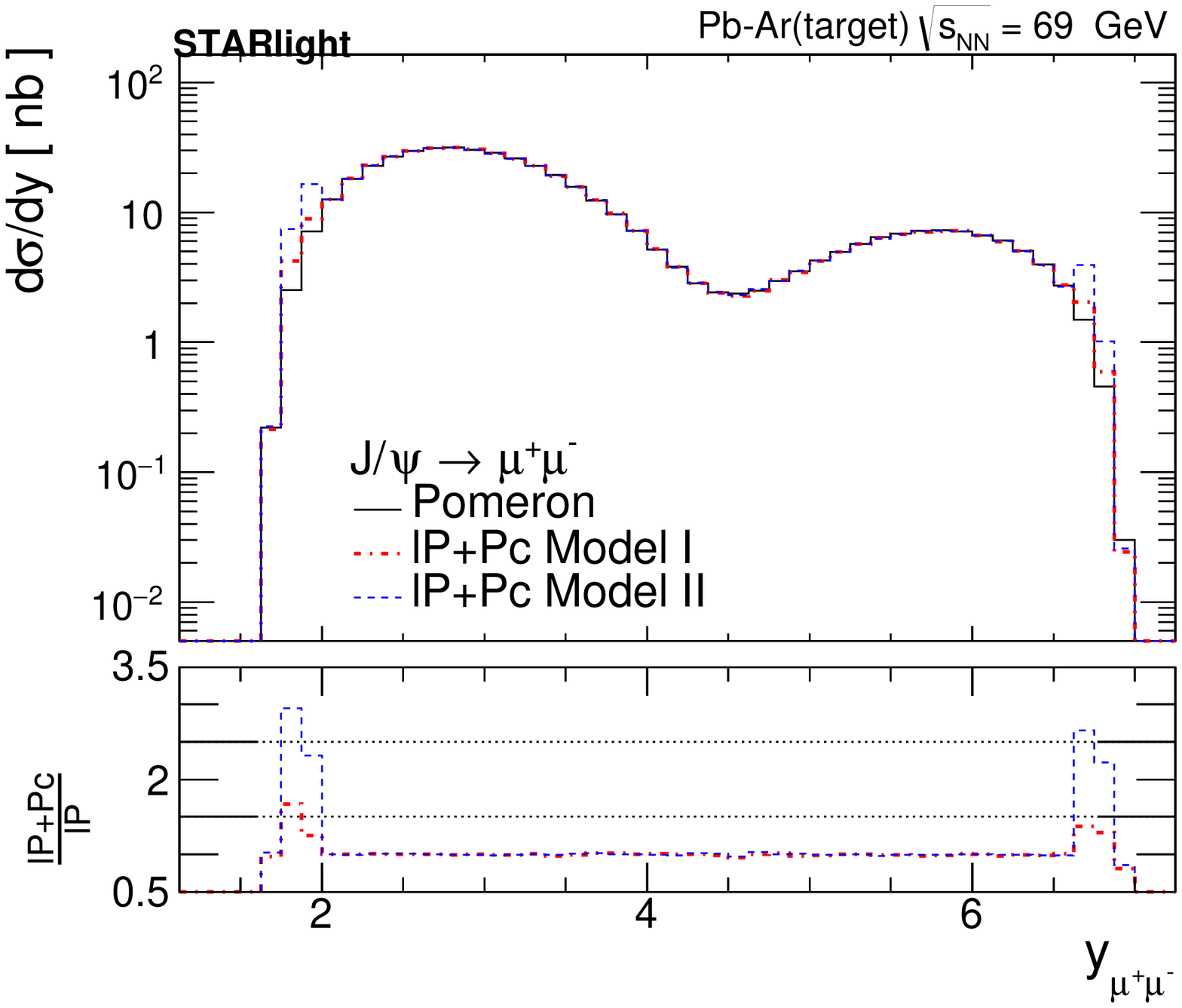} & \hspace{0.2cm} &
\includegraphics[scale=0.45]{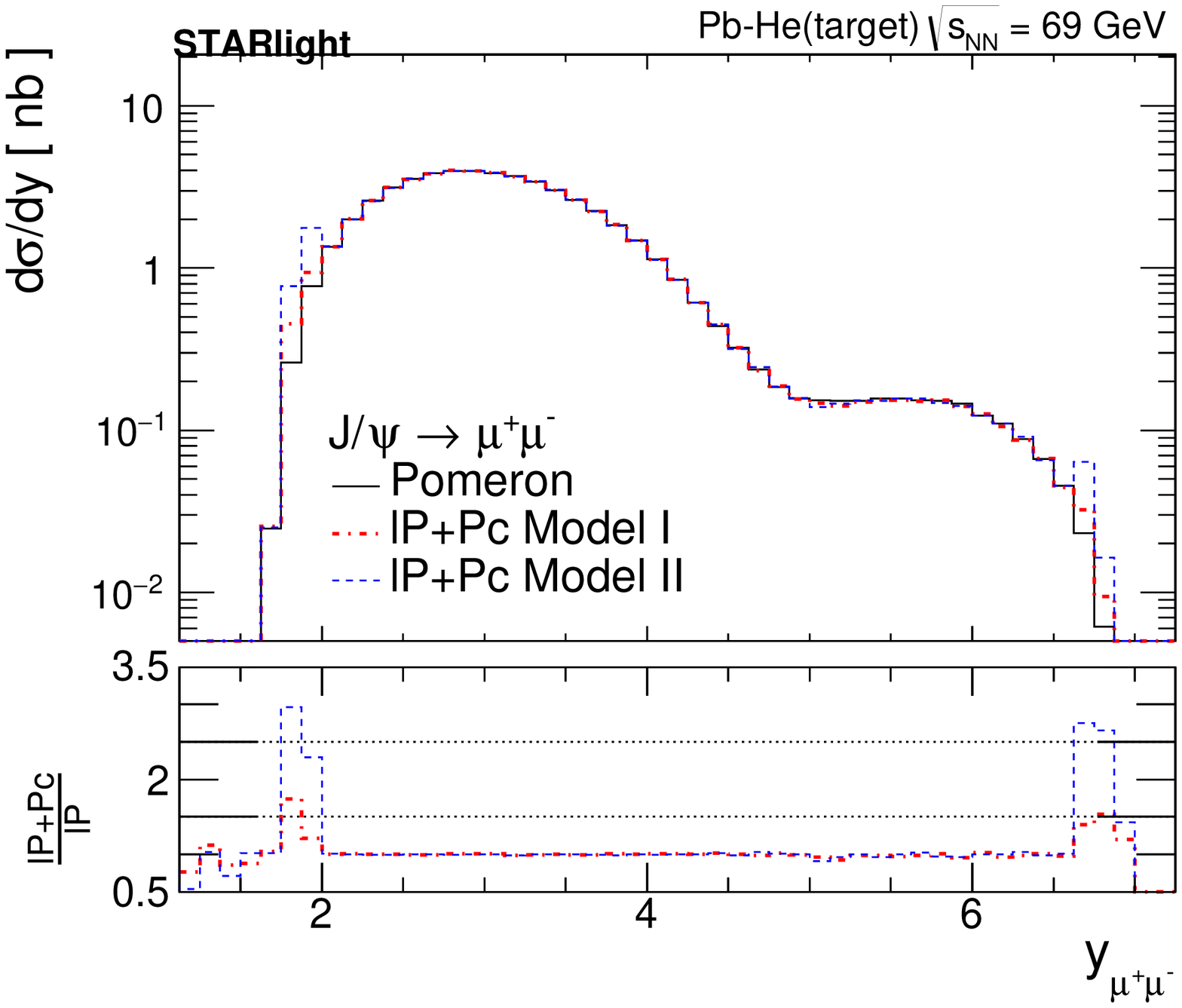} \\
(a) & \,\,\, & (b)
\end{tabular}
\caption{Rapidity distribution for the $J/\Psi$ photoproduction in (a) $Pb \,Ar $  and (b) $Pb \, He$  fixed -- target collisions  at LHC. }
\label{Fig:penta2}
\end{figure}

As the recent studies were performed by injecting the  noble gases ($He, \, Ne, \, Ar$) in the LHC beam pipe and   similar configurations  should also be present in future analysis, we  will extend our previous analysis for the $J/\Psi$ photoproduction in $Pb\,Ar$ and $Pb\,He$ fixed -- target collisions. The predictions are presented in Fig. \ref{Fig:penta2}. As in the case of $pPb$ fixed -- target collisions, we predict an enhancement of the rapidity distribution for $y_{\mu^+ \mu^-} \approx  \, 1.8$ that is larger for the Model II than for the Model I. We also predict a second enhancement of the distribution for 
$y_{\mu^+ \mu^-} \approx  \, 6.7$ which is associated to events associated to photon emitted by the target. Unfortunately, the position of this second enhancement is beyond the rapidity range covered by the current detectors.

 As a summary, in this paper we have investigated the impact of the $P_c$ resonances on the $J/\Psi$ photoproduction at the RHIC and LHC.  
  We have considered $pA$ collisions at the RHIC and LHC, as well $pPb$, $Pb \, Ar$ and $Pb \,He$ fixed -- target collisions at the LHC, and estimated the rapidity distributions of the $J/\Psi$ meson taking into account its decay into a $\mu^+ \mu^-$ pair. Assuming two models for the description of the $P_c$ production, we have derived upper limits for the production of different $P_c$ states at the LHC. Our goal was to verify if the study of this process can be useful to confirm  the existence of the resonances as well to constrain its properties. We shown that the presence of the resonances modifies the associated rapidity distribution due to the large enhancement of the $\gamma p \rightarrow J/\Psi p$ cross section near the threshold. We have demonstrated that in the collider mode, the rapidity distribution is enhanced for  
 $y_{\mu^+ \mu^-} \approx - \, 7.5$, which is beyond of rapidity range covered by the current detectors of the LHC. On the other hand, in the fixed -- target mode, this enhancement is predicted to occur in a rapidity range that can be covered in future fixed -- target studies.
 Our results indicate that the study of $\gamma p$ interactions
at LHC can also provide complementary and independent checks on the properties of the  pentaquark states, and help to understand their underlying nature.

\section*{Acknowledgments} 
VPG acknowledges useful discussions with S. R. Klein and would like to express a special thanks to the Mainz Institute for Theoretical Physics (MITP) of the Cluster of Excellence PRISMA+ (Project ID 39083149) for its hospitality and support.   
This work was partially financed by the Brazilian funding agencies CAPES, CNPq,  FAPERGS and INCT-FNA (process number 464898/2014-5).

\end{document}